\documentclass[11pt]{article}
\usepackage{epsfig}
\usepackage{amsmath}
\usepackage{amssymb}
\usepackage{mathrsfs}
\usepackage{graphicx}
\newsavebox{\PSLASH}
\sbox{\PSLASH}{$p$\hspace{-1.8mm}/}
\newcommand{\PS}{\usebox{\PSLASH}}
\newsavebox{\KSLASH}
\sbox{\KSLASH}{$k$\hspace{-1.8mm}/}
\newcommand{\KS}{\usebox{\KSLASH}}

\begin{document}

\vspace{4cm}
\begin{center}
{\Large\bf{Top Quark Forward-Backward Asymmetry and $W'$-Boson with
    General Couplings }}\\

\vspace{1cm}
{\bf Seyed Yaser Ayazi, Sara Khatibi, and Mojtaba Mohammadi Najafabadi }\\
\vspace{0.5cm}
{\sl  School of Particles and Accelerators, \\
Institute for Research in Fundamental Sciences (IPM) \\
P.O. Box 19395-5531, Tehran, Iran}\\

\vspace{2cm}
 \textbf{Abstract}\\
 \end{center}
The measured forward-backward asymmetry in top pair events at the Fermilab Tevatron collider deviates significantly
from the standard model expectation. Several models have been proposed to describe
the observed asymmetry which grows with the rapidity
difference of the top pairs and also with the $t\bar{t}$ invariant mass.
The presence of a heavy charged gauge boson $W'$ with the coupling $W'-t-d$ (left-handed, right-handed, and
a mixture of left and right-handed couplings)
could generate the desired top forward-backward asymmetry keeping the top pair cross section consistent with
the standard model prediction.
Such $W'$-boson makes contribution to the electric dipole moment of the neutron through contribution
to the $d$-quark electric dipole moment, recently measured charge asymmetry ($A_{C}$)
by the LHC experiments, and the total cross section of top pair at the LHC and Tevatron. We show that the upper bounds
on neutron and top electric dipole moments disfavour any $W'$ with a mass below 240 GeV which could explain the
Tevatron forward-backward asymmetry. It is shown that the charge asymmetry
provides an allowed region in the parameters space with no overlap with the allowed region where the asymmetry could be described.
\\

PACS number(s): 14.65.Ha, 12.60.-i

\newpage

\section{Introduction}

Examination of the top quark interactions with other particles offer a window to possible new physics
beyond the standard model (SM).
Top quark with a mass close to the vacuum expectation value $v \simeq$ 246 GeV,
seems to be more sensitive to the electroweak symmetry breaking mechanism than other
standard model particles.
At hadron colliders, top quarks are produced either singly via electroweak
interactions or in pair via strong interactions with larger cross section.
Many $t\bar{t}$ events have been produced at Tevatron and LHC up to now
and their experiments provide the possibility to study the top quark
properties \cite{werner},\cite{beneke},\cite{werner2}.

So far, the only observed inconsistency with the standard model
predictions has been the forward-backward
asymmetry in top pair production at the Tevatron.
The top quark forward-backward asymmetry is defined as
the difference of the number of events with $\cos\theta > 0$
and $\cos\theta < 0$, where $\theta$ is the top quark production angle in the $t\bar{t}$ rest frame:
\begin{eqnarray}\label{afb}
A_{FB} = \frac{N_{t}(\cos\theta > 0)-N_{t}(\cos\theta < 0)}{N_{t}(\cos\theta > 0)+N_{t}(\cos\theta < 0)}
\end{eqnarray}

The measured $A_{FB}$ from
the CDF and D0 experiments are $A_{FB} = 0.158 \pm 0.075$ \cite{cdfafb},
$A_{FB} = 0.196 \pm 0.065$ \cite{d0afb}.
These measurements do not agree with the SM expectation, 0.089 \cite{afbsm1},\cite{afbsm2},\cite{afbsm3},\cite{afbsm4}.
Since the sign of $\cos\theta$ and $y_{t}-y_{\bar{t}}$ are the same, the forward-backward asymmetry could be defined through the difference between the rapidities 
of top and anti-top quarks according to the following relation:
\begin{eqnarray}
A_{FB} = \frac{N_{t}(y_{t}>y_{\bar{t}})-N_{t}(y_{t}<y_{\bar{t}})}{N_{t}(y_{t}>y_{\bar{t}})-N_{t}(y_{t}<y_{\bar{t}})}
\end{eqnarray}

An interesting observation is that the CDF Collaboration measured
the asymmetry as a function of the invariant mass of $t\bar{t}$ and found that the deviations of asymmetry
grow with $m_{t\bar{t}}$ \cite{cdfnote}.
While measurements of the differential
cross section of $d\sigma/dm_{t\bar{t}}$ show a good agreement with the
SM expectations and therefore no evidence of beyond SM
physics in $m_{t\bar{t}}$ spectrum.
Accordingly, one should note that any new physics which explains the $t\bar{t}$ forward-backward asymmetry
must satisfy all other measurements consistent with the SM predictions.
There have been several models proposed to describe the observed asymmetry by Tevatron
experiments \cite{fb1},\cite{fb2},\cite{fb3},\cite{fb4},\cite{fb5},\cite{a1},\cite{a2},\cite{a3},\cite{a4},\cite{a5}.

One of the models
is proposing a new heavy charged gauge boson with the following coupling to the top quark \cite{wp1}:
\begin{eqnarray}\label{lag}
\mathcal{L} = -g'W'^{+}_{\mu}\bar{t}\gamma^{\mu}(g_{L}P_{L}+g_{R}P_{R})d+h.c.
\end{eqnarray}
where $g'$ is the coupling constant, $g_{R,L}$ are the chiral couplings
of the $W'$ boson with fermions,$P_{L,R} = (1\pm \gamma_{5})/2$ are the chirality projection operators.
Recent LHC data excluded the existence of any SM-like $W'$-boson ($g'=g,g_{L}=1,g_{R}=0,$)
by direct measurements into final states of leptons below the mass of
2.15 TeV \cite{wprime}. Therefore, we do not consider a SM-like $W'$-boson in this analysis.

The new Lagrangian involving $W'$ interactions with the top quark and $d$-quark leads to new diagram for
top pair production at the Tevatron and LHC through the subprocess $d\bar{d}\rightarrow t\bar{t}$.
Hence, it contributes to the differential and total cross sections of $t\bar{t}$ at the LHC and Tevatron.

In addition to the $t\bar{t}$ production cross section, the $W'$-boson
introduced in the new Lagrangian in Eq.\ref{lag},
can induce new contribution to the $d$-quark electric dipole moment.

In this letter, three categories of this model according to different values of $g_{L,R}$ are considered:
(1) $g_{L} = g_{R} = 1$ ($g'$ and $M_{W'}$ are free parameters). (2) $g_{L} = 0,g_{R} = 1$ ($g'$ and $M_{W'}$ are free parameters).
(3) $g'$ is removed and in general $g_{L}, g_{R}$ and $M_{W'}$ are free parameters.
Then, we try to find allowed regions in the parameters space consistent with the top pair cross section
at the LHC, Tevatron, the forward-backward asymmetry, upper limits on the top and $d$-quark electric dipole moment, and
finally the charge asymmetry in $t\bar{t}$ events.

The paper is organized as follows. The next section is dedicated to describe the new
physics corrections to the top pair production cross section. In section 3, we describe
the role of new Lagrangian in the electric dipole moments of $d$-quark and top quark.
And finally, a numerical calculations and discussion on the results are presented in section 4.

\section{Cross Sections and Charge Asymmetry}

As mentioned in the previous section, in addition to the $s$-channel
diagram from the gluon exchange (SM contribution) in $d\bar{d}\rightarrow t\bar{t}$ process,
a $t$-channel diagram due to $W'$ exchange has to be added in the calculations.
Therefore, the squared matrix element ( after ignoring the $d$-quark mass, summing over the spin and color, and averaging over the
color and spin of the initial partons)
has the following form \cite{wp1},\cite{wp2}:
\begin{eqnarray}
\overline{|\mathcal{M}|^{2}} = \frac{4g_{s}^{4}}{9\hat{s}^{2}}(u_{t}^{2}+t_{t}^{2}+2\hat{s}m_{t}^{2})+
\frac{4g'^{2}g_{s}^{2}}{9\hat{s}t_{W'}}(g_{L}^{2}+g_{R}^{2})[2u_{t}^{2}+2\hat{s}m_{t}^{2}+
\frac{m_{t}^{2}}{M_{W'}^{2}}(t_{t}^{2}+\hat{s}m_{t}^{2})] + \\ \nonumber
\frac{g'^{4}}{4t_{W'}^{2}}[4((g_{L}^{4}+g_{R}^{4})u_{t}^{2}+ 2g_{L}^{2}g_{R}^{2}\hat{s}(\hat{s}-2m_{t}^{2})) +
\frac{m_{t}^{4}}{M_{W'}^{4}} (g_{L}^{2}+g_{R}^{2})^{2}(t_{t}^{2}+4M_{W'}^{2}\hat{s})]
\end{eqnarray}
where $\hat{s},t,u$ are mandelstam parameters, and $u_{t} = u - m_{t}^{2}, t_{t} = t- m_{t}^{2}, t_{W'}=t-M_{W'}^{2}$.
The first term in the above amplitude is the SM contribution, the second term is the interference term between
the standard model and the new diagram arising from $W'$-exchange, and
the last term is the contribution of $W'$-boson exchange.

The total and differential cross sections at hadron colliders can be obtained by convoluting the partonic
cross section with the parton distribution functions (PDF) for the initial hadrons.
Using CTEQ6L \cite{cteq} set as the parton distribution functions, the total cross section at the Tevatron and LHC
is calculated. The top quark forward-backward asymmetry is also calculated according to the Eq.\ref{afb}.

The total measure production cross section of the $t\bar{t}$ at the Tevatron and the LHC are \cite{cstev},\cite{cslhc}:
\begin{eqnarray}
\sigma_{\text{Tevatron}} = 7.56\pm0.63~\text{pb}~,~\sigma_{\text{LHC}} = 165\pm 13.3~\text{pb}.
\end{eqnarray}

At the LHC, since the initial state is symmetric (proton-proton), the top quark forward-backward asymmetry vanishes.
However, an asymmetry in charge $A_{C}$ can be measured, which is defined as
the relative difference between $t\bar{t}$ events with $|y_{t}| > |y_{\bar{t}}|$ and $|y_{t}| < |y_{\bar{t}}|$.
Where $y_{t} (y_{\bar{t}})$ are the rapidity of the top
(anti)quark in the laboratory frame.
In proton-proton collisions at the LHC,
the $u, d$ valence quarks carry larger average momentum
fraction than the anti-quarks. This leads
to a boost of the $t\bar{t}$ system along the direction of the
\textit{incoming quark}, and therefore to a larger average rapidity for
top quarks than anti-top quarks.
The ATLAS and CMS measurements
for the charge asymmetry are: $A_{C} = -0.018 \pm 0.036$ \cite{atlasac}, $A_{C} = -0.013 \pm 0.041$ \cite{cmsac}
, and the SM prediction is $A_{C} = 0.0115$ \cite{afbsm4}.
It is notable that within the uncertainties the standard model prediction for charge asymmetry
is in agreement with the measured values at the LHC.
From another side, since there is some tension between the asymmetry in charge at the LHC and 
the forward-backward asymmetry at the Tevatron, it was expected to observe an enhancement 
of $A_{C}$ at the LHC \cite{juan1},\cite{juan2}.

We probe the parameters space of considered model in Eq.\ref{lag} with the current measured $t\bar{t}$ production cross section both
at the Tevatron and the LHC, the forward-backward asymmetry, and the charge asymmetry of top pair events
at the LHC.

\section{Electric Dipole Moment Analysis}

The electric dipole moment of a spin $1/2$ particle is defined by the effective Lagrangian \cite{cp1}:
\begin{eqnarray}
\mathcal{L} = -\frac{i}{2}d_{f}\bar{\psi}\sigma_{\mu\nu}\gamma_{5}\psi F^{\mu\nu}
\end{eqnarray}

where $d_{f}$ stands for the fermion $f$ electric dipole moment.
We notice that this Lagrangian is CP violating while the common
standard model term which describes the interaction of a fermion
with photon ($-iQe\gamma_{\mu}$) is CP conserving. More information about
electric dipole moment and CP violation effects can be found in
several papers such as \cite{cp1},\cite{cp2},\cite{cp3}.

\begin{figure}
\centering
  \includegraphics[width=10cm,height=7cm]{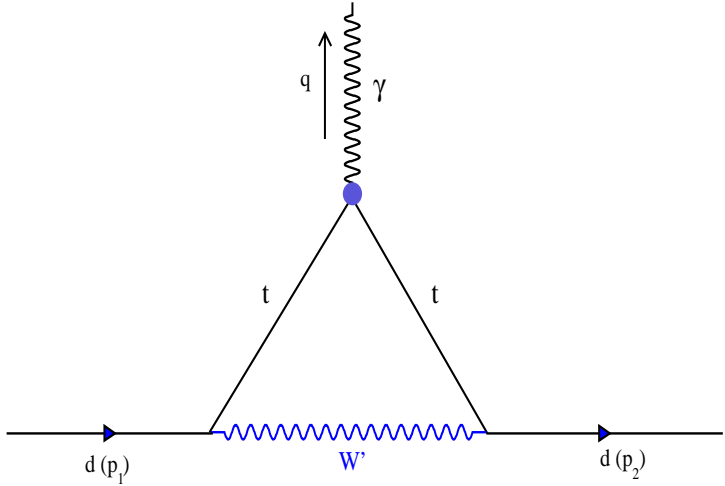}
  \caption{\textsl{Diagram contributing to the on-shell $q\bar{q}\gamma$
 vertex arising from the $W'$ interactions.} }\label{loop1}
\end{figure}

The contribution of the new couplings of $W'-t-d$ to the on-shell $q\bar{q}\gamma$
coupling is given by the diagram shown in Fig. \ref{loop1}. The respective one-loop vertex is:
\begin{eqnarray}
\Gamma_{\mu}=-g'^{2}d_t\int{\frac{d^{4}k}{(2\pi)^{4}}\frac{g^{\alpha\beta}\gamma_{\alpha}(g_{L}P_{L}+g_{R}P_{R})
(\KS-\PS_{2}+m_{t})\gamma_{5}\sigma_{\mu\nu}q^{\nu}(\KS-\PS_{1}+m_{t})
\gamma_{\beta}(g_{L}P_{L}+g_{R}P_{R})}{(k^2-M_{W^{\prime}}^2)((k-p_1)^2-m_t^2)((k-p_2)^2-m_t^2)}}
\end{eqnarray}

where $k$ is the momentum of the $W'$-boson, $p_{1,2}$ are the momenta of
the $d$-quark as depicted in Fig. \ref{loop1}.
There are contributions to electric and magnetic dipole moments of
the $d$-quark, however, we are only interested in CP violating terms.
After some algebraic manipulations, using Gordon identity and Dirac equation
and finally integration over $k$, we obtain:
\begin{eqnarray}\label{dt}
d_{d} = \frac{g'^{2} d_{t}}{ 16\pi^{2}} [2x_{d}x_{t}(g_{L}^{2}+g_{R}^{2})+16x_{d}^{2}g_{R}g_{L}]f(x_{t},x_{d})
\end{eqnarray}
where $x_{t} = m_{t}^{2}/M_{W'}^{2}$, $x_{d} = m_{d}^{2}/M_{W'}^{2}$ and
\begin{eqnarray}\label{int}
f(x_{t},x_{d}) = \int^{1}_{0} dx \int_{0}^{1-x} dy \frac{1-x-y}{1-(1-x_{t})(x+y)-x_{d}(1-x-y)(x+y)}
\end{eqnarray}
In the denominator of the integrand in Eq.\ref{int}, the third term is
neglected because  
$x_{d}$ is negligible with respect to $x_{t}$ $(x_{d}/x_{t}\thicksim 10^{-9})$.  Furthermore,
we will see that the second term of Eq.\ref{dt} does not affect the results due to the 
same reason.
The two-dimensional integral can be approximated as:
\begin{eqnarray}\label{fx}
f(x_{t}) = \frac{x_{t}^{2}-1-2x_{t}\log x_{t}}{2(x_{t}-1)^{3}}
\end{eqnarray}

At $x_{t} = 1$ or $M_{W'} = m_{t}$, this function is indeterminate and it approaches $\frac{1}{6}$ when $x_{t}\rightarrow 1$.
However, as discussed in \cite{wp1}, if the $W'$ is light enough the top quark can decay into a $d$-quark 
and a $W'$-boson and definitely, it would have been already observed in the top quark decays at the Tevatron. Therefore,
in this work we concentrate on the $W'$ boson with a masss higher than 200 GeV.

Now, we use one of the most used approach to predict the quark
electric dipole moment contribution
to the neutron electric dipole moment. This
originates from the $SU(6)$ quark model, with assuming a non-relativistic
wave-function to the neutron.
The neutron electric dipole moment has the following form in terms of
quarks electric dipole moments \cite{cp1}:
\begin{eqnarray}
d_{n} = \eta (\frac{4}{3}d_{d}-\frac{1}{3}d_{u})
\end{eqnarray}
where $d_{d},d_{u}$ are the down and up quark electric dipole moments,
respectively. The QCD correction factor $\eta$ is around 0.61.

The current experimental upper limit on the neutron electric dipole moment is
$d_{n} < 2.9\times 10^{-26}$ e.cm. \cite{nedm1},\cite{nedm2}.
Future experiments are able to measure the neutron electric dipole moment
down to $10^{-28}$ e.cm. Therefore, in addition to current limit, we
also present the results with the future upper bound \cite{sns}.

Applying the upper
limits on the top quark and neutron electric dipole moments in Eq.\ref{dt}
provides consistent region in the $(g',M_{W'})$ plane with these values.

The upper limit on top quark electric dipole moment has been
extracted from the upper limit on the branching ratio of
$b\rightarrow s\gamma$. It has been found to be less than $10^{-16}$ e.cm. \cite{topedm1},\cite{topedm2}.

\section{Numerical Results and Discussion}

In the numerical calculations, the top quark mass has been set $m_{t} = 172.5$ GeV.
The contributions of the new physics with $W'-d-t$ to the observables defined in section 2
at the partonic level is calculated  by employing CTEQ6L parton distribution functions \cite{cteq}.
The calculation is performed at fixed renormalization and factorization scale $\mu_{R} = \mu_{F} = m_{t}$.
To include the effect of higher order QCD corrections to the observables, all observables are normalized to
ratio of measured experimental cross section to the leading order SM cross section.
The major QCD corrections within the $W'$ model have only been computed in \cite{wp2} and leaded to almost the same as pure
SM higher order QCD corrections.

In order to obtain the relevant parameters of the $W'$
model, we scan the region $200 < M_{W'} < 800$ GeV in three different
categories.

\begin{figure}
\centering
  \includegraphics[width=12cm,height=8cm]{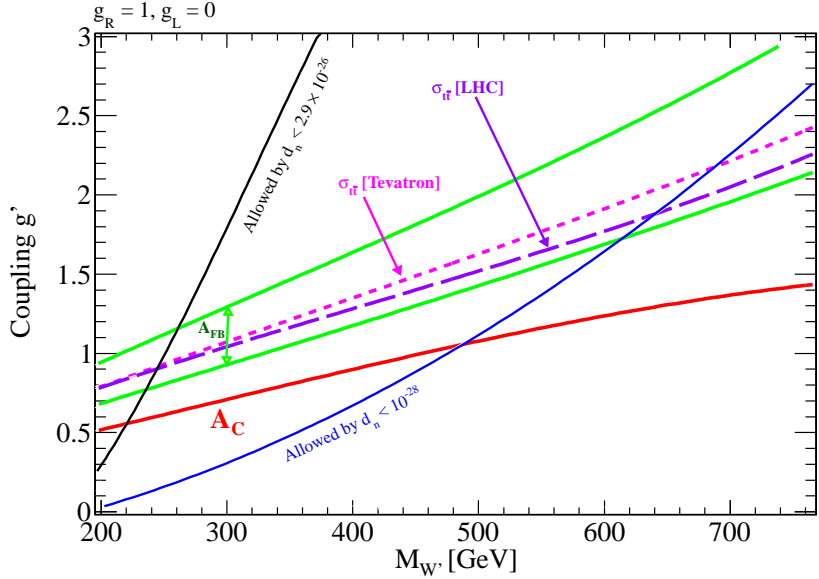}
  \caption{\textsl{Region of $W'$ coupling in terms of $M_{W'}$ consistent
with Tevatron measurements of the $t\bar{t}$ forward-backward asymmetry (region between two green curves).
The areas between the $M_{W'}$-axis and the dashed curves are consistent with the LHC and Tevatron cross sections. The consistent
region with the LHC charge asymmetry is the area between the solid red curve and $M_{W'}$-axis. The 
area under the black curve is the allowed region coming from limits on electric dipole
moments. The region under the blue curve is corresponding to the
allowed region with $10^{-28}$ as for the neutron electric dipole moment.}}\label{gr}
\end{figure}

$\bullet{~g_{L} = 0, g_{R} = 1}$: The area of $g'$ coupling versus $M_{W'}$ consistent
with Tevatron measurement of the $A_{FB}$ asymmetry is the band between two green curves in Fig.\ref{gr}.
The regions between the $M_{W'}$-axis and the dashed lines are allowed regions which are consistent
with the LHC and Tevatron top pair cross section measurements. The consistent region with the
LHC measurement of charge asymmetry is the region between the red solid curve and the $M_{W'}$-axis.
The area under the black (blue) curve is the allowed region in the $M_{W'},g'$ plane that arises from the electric dipole moment
bound of $10^{-26}$ ($10^{-28}$) e.cm.

As it can be seen in Fig. \ref{gr}, the acceptable values for $g_{R}$ increases slightly with the mass of $W'$.
The electric dipole moment excludes any right-handed $W'$ with the mass below 240 GeV which could explain the
forward-backward asymmetry. Since there is no overlapping region between the measured charge asymmetry at the LHC and $A_{FB}$ area,
$A_{C}$ disfavors right-handed $W'$-boson which could explain the top quark forward-backward
asymmetry.

Other interesting point depicted in Fig. \ref{gr} is the allowed values of $g'$ and $M_{W'}$
arising from the future upper limit on the neutron electric dipole moment.
Future experiments are able to measure the electric dipole moment of neutron down to $10^{-28}$ e.cm., i.e. two
orders of magnitude better than the present limits \cite{sns}. In
Fig.\ref{gr}, the light blue points are the allowed
region according to the future bound on the neutron electric dipole
moment. As it can be seen neutron electric dipole moment would be able
to exclude the existence of any right-handed $W'$-boson with a mass
below 600 GeV which could explain the forward-backward asymmetry.
As a consequence of the behaviour of $f(x)$ in Eq.\ref{fx} at $x=1$,
some points are allowed at $M_{W'}=m_{t}$ which can be seen in Fig.\ref{gr}.

\begin{figure}
\centering
  \includegraphics[width=12cm,height=8cm]{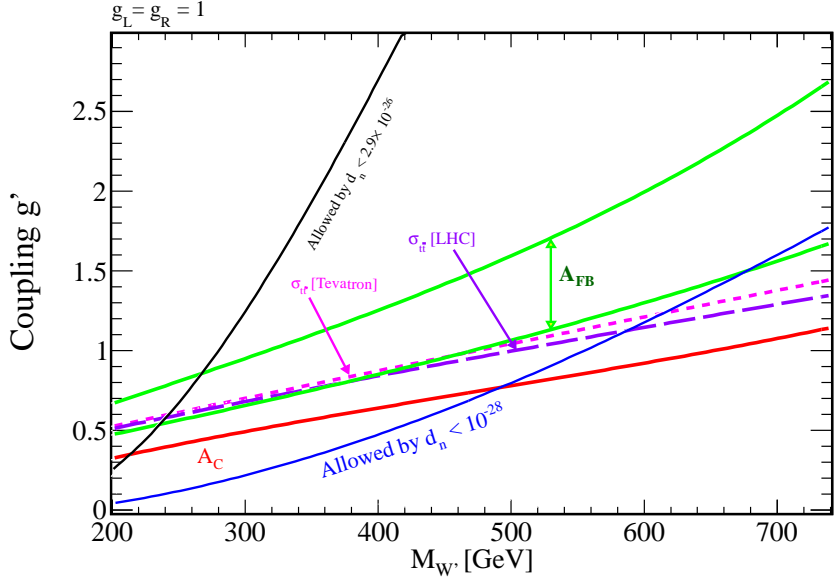}
  \caption{\textsl{Region of $W'$ coupling with $g_{L} = g_{R} = 1$ versus $M_{W'}$ consistent
with Tevatron measurements of the $t\bar{t}$ forward-backward asymmetry (region between two green curves).
The areas between the $M_{W'}$-axis and the dashed curves are consistent with the LHC and Tevatron cross sections. The consistent
region with the LHC charge asymmetry is the area between the solid red curve and $M_{W'}$-axis. The area
under the black curve depicts the allowed region arises from bounds on electric dipole
moments. The region under the blue curve is corresponding to the
allowed region with $10^{-28}$ e.cm as for the neutron electric dipole moment.}}\label{grgl}
\end{figure}

$\bullet{~g_{L} = g_{R} = 1}$: As depicted in Fig. \ref{grgl}, the consistent area in the plane of $g'$ coupling and $M_{W'}$
with Tevatron measurement of the $A_{FB}$ asymmetry (the band between two green curves) and the LHC and Tevatron top pair cross
section measurements is small.
In other words, only low mass vector like $W'$ can satisfy the values of $A_{FB}$ and
cross sections.
The regions between $M_{W'}$-axis and the dashed lines are allowed regions consistent
with the LHC and Tevatron top pair cross section measurements.
Only the region of $M_{W'}$ below 320 GeV can explain the asymmetry. While the current limit on electric dipole moment of neutron
disfavors the $W'$-boson mass up to 240 GeV. Similar to the latter case, the charge asymmetry measurement does not
have any overlapping region with the favorite region. Combining the electric dipole moment limit 
with the cross section limits, we find that only $W'$ boson with $240 < M_{W'} < 320$ GeV is consistent 
with the Tevatron $A_{FB}$.

The same as previous case, as shown in Fig. \ref{grgl}, the light blue dots are the allowed
region according to the future bound on the neutron electric dipole moment. Clearly, the future limit on neutron electric
dipole moment is able to exclude a vast region in the ($g',M_{W'}$) plane which could explain the forward-backward asymmetry.
Therefore, this case of $W'$ coupling is not
able to describe the asymmetry because of the cross sections limits.

\begin{figure}
\centering
  \includegraphics[width=7cm,height=6cm]{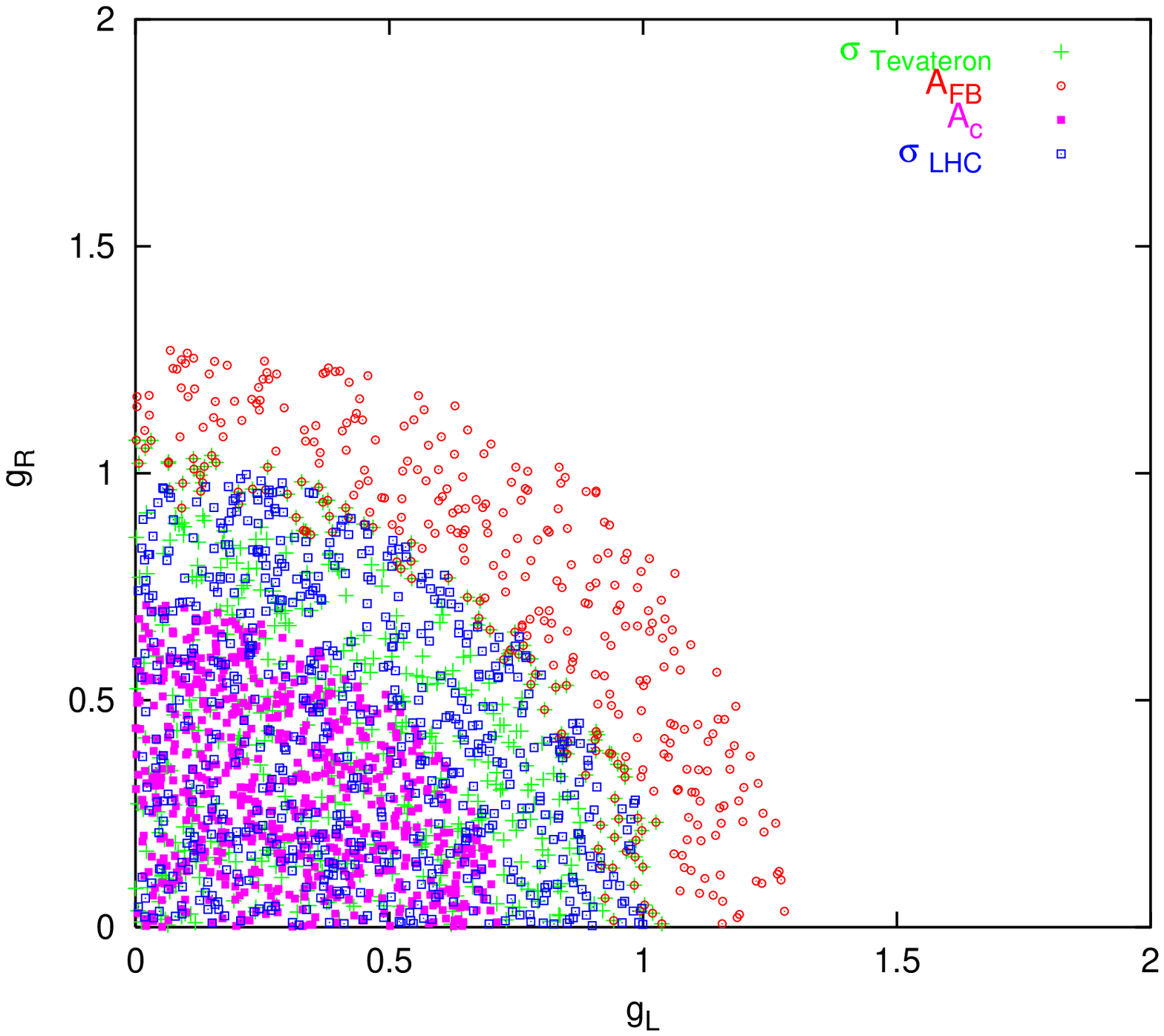}
  \includegraphics[width=7cm,height=6cm]{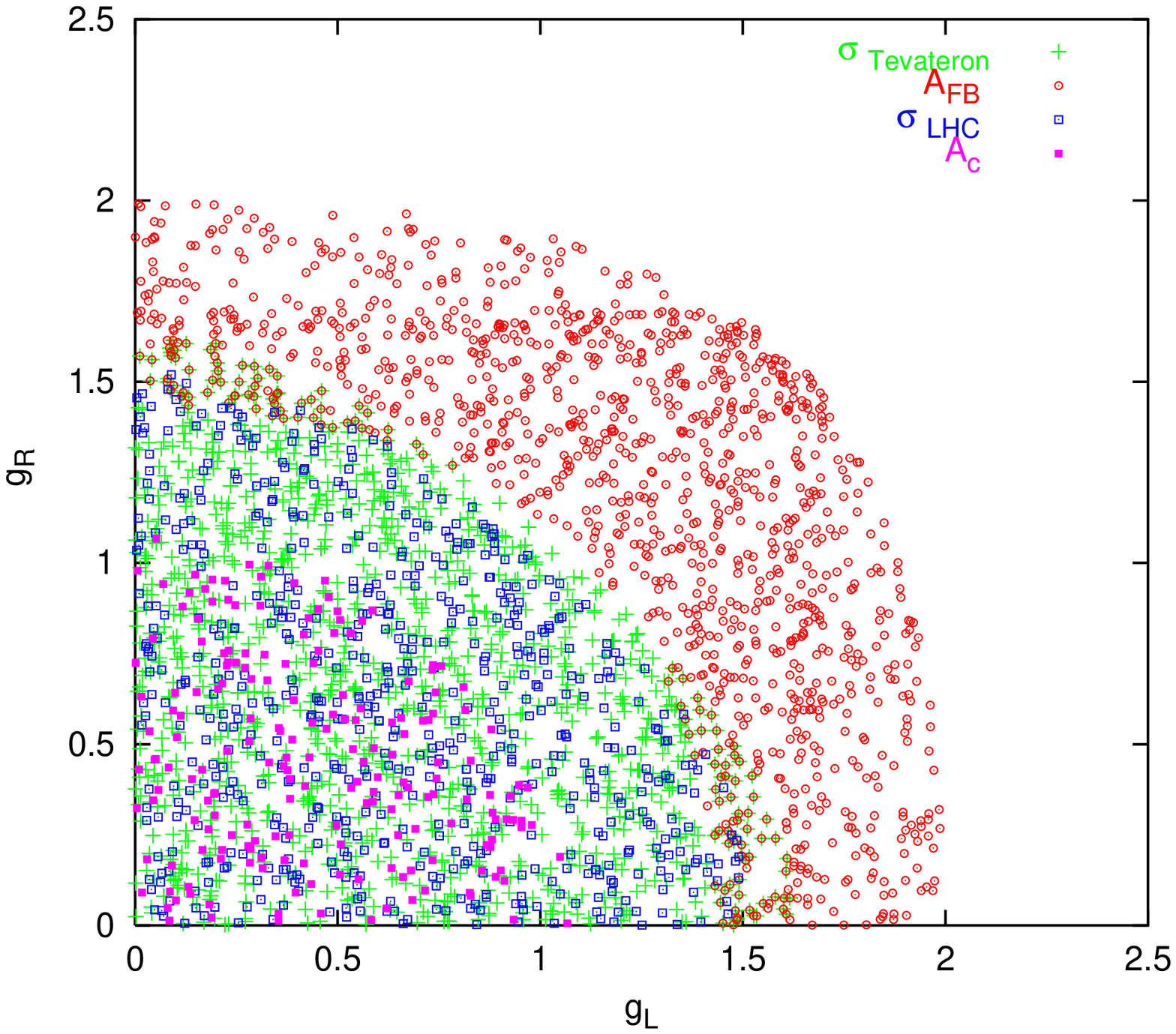}\\
  \includegraphics[width=7cm,height=6cm]{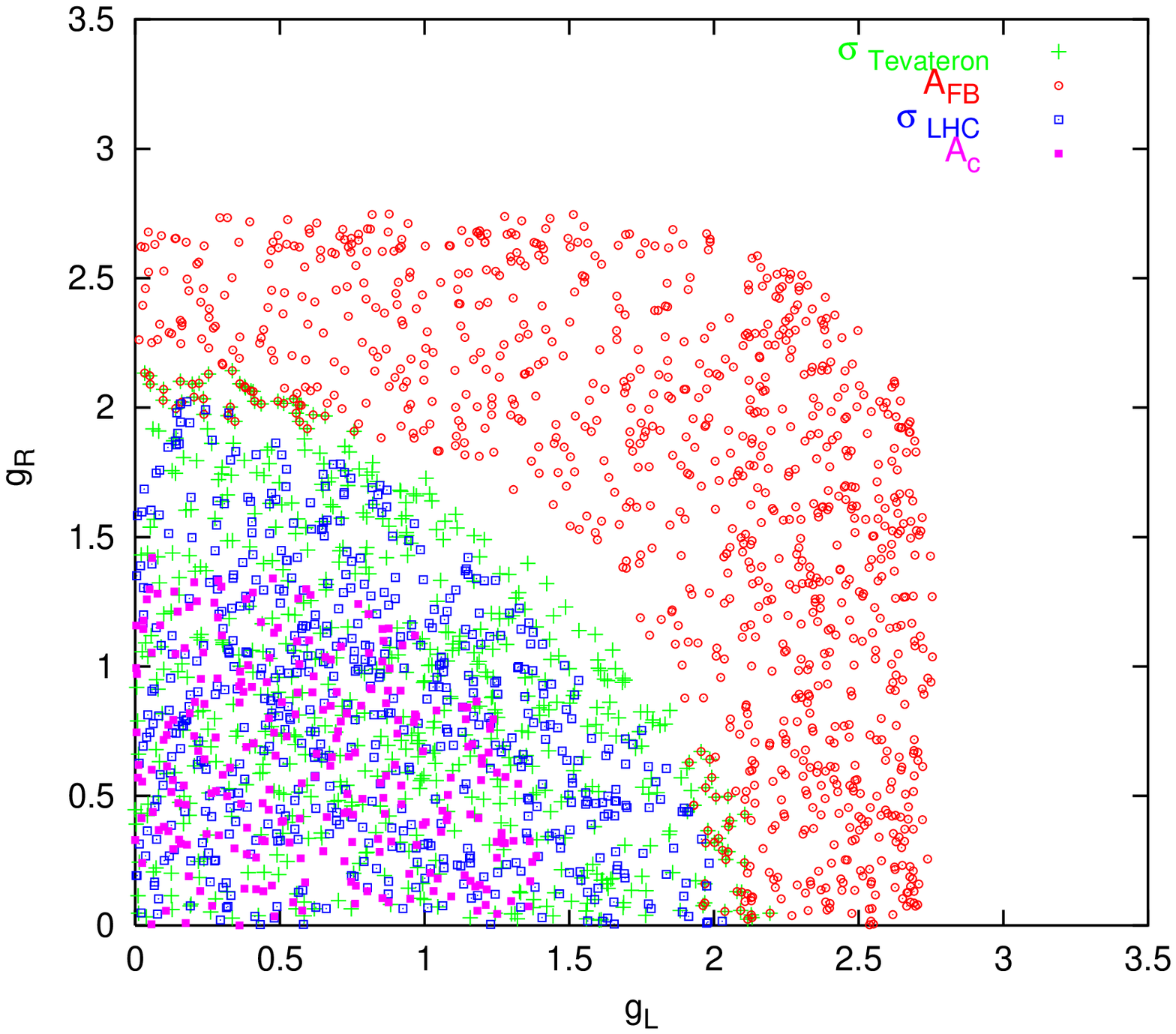}
  \caption{\textsl{The allowed regions consistent with the Tevatron forward-backward asymmetry, $t\bar{t}$ cross sections at the
LHC and Tevatron, the LHC charge asymmetry with general values of
$g_{R},g_{L}$ couplings. In the left top plot $M_{W'} = 300$ GeV and
for the right top plot $M_{W'}$ is equal 500 GeV, in the bottom plot
$M_{W'}$ = 700 GeV.}}\label{mw300}
\end{figure}

$\bullet{~g' = 1, \text{arbitrary}~g_{L},g_{R}}$: In this case, there are three parameters $g_{L},g_{R}$ and $M_{W'}$ which could
be found in such a way that the top forward-backward asymmetry be generated. Figs.\ref{mw300} show the regions
consistent with the measured top pair cross sections at the Tevatron (depicted by plus symbol), LHC (empty squares), and the Tevatron $t\bar{t}$
forward-backward asymmetry (empty circles) for $M_{W'} =
300~\text{(left top)},500~\text{(right top)},~700~ GeV \text{(bottom)}$. We note that the electric dipole moment
bounds do not provide comparable results with the other measurements.
Comparison of plots for $M_{W'} = 300$ GeV and $M_{W'} = 500, 700$ GeV leads to the fact that
the overlapping region between the measured cross sections and the Tevatron forward-backward asymmetry
gets smaller with increasing the $W'$-boson mass.
According to Figs. \ref{mw300}, with growing the $W'$-boson mass
 the cross sections and forward-backward asymmetry are
both satisfied by either large $g_{L}$ or large $g_{R}$. It means that
in the $(g_{L},g_{R})$ plane, the regions with intermediate values of $g_{R}$ and $g_{L}$ are
excluded. It also could be seen in Fig. \ref{grgl}, with increasing
$M_{W'}$ the overlapping area decreases because of different slopes of
cross sections  and $A_{FB}$ lower limit curve.
However, in analogy with the previous cases, no overlapping region between the LHC charge asymmetry (pink squares) and the Tevatron top forward-backward asymmetry
is seen.

\section{Conclusions}

In this paper, we investigated the consistency of heavy charged vector boson $W'$ with
the existing $t\bar{t}$ production measurements. Various cases of $W'$ couplings with top and down
quarks are examined: right-handed $W'$ ($g_{L}=0,g_{R}=1$), vector-like ($g_{L}= g_{R}=1$), and in general mixture of left and right-handed
(arbitrary $g_{L,R}$).
We find no overlapping region between the large positive $t\bar{t}$ forward-backward asymmetry measurements at the Tevatron and the
LHC charge asymmetry measurements in all cases.
Apart from the LHC charge asymmetry, the electric dipole moment limits do not support a $W'$ boson below
240 GeV which could explain the Tevatron forward-backward asymmetry.
We also found that the future upper limit on the neutron electric dipole moment is not in favor of 
any right-handed $W'$-boson with a mass under around 600 GeV which could explain the top quark forward-backward asymmetry.

$\bullet$ {\bf Note added}
At around the same time as our paper was being written a related analysis
of the charge asymmetry part with right-handed coupling $W'$ appeared in \cite{Fajfer:2012si}.


\end{document}